\documentclass[manuscript,preprintnumbers,prb,aps,amssymb,superscriptaddress]{revtex4}
\usepackage{graphicx}% Include figure files
\usepackage{dcolumn}% Align table columns on decimal point
\usepackage{bm}% bold math

\begin{document}

%\title{Growth and characterization of single crystalline SrZr$_{x}$Ti$_{1-x}$O$_3$ $x$ = 0.7 on Ge(100)}% Force line breaks with \\

\title{Band-gap engineering at a semiconductor - crystalline oxide interface}% Force line breaks with \\

\author{J. Moghadam}
 \affiliation{Department of Physics, The University of Texas at Arlington, 502 Yates Street, Arlington, TX  76019, USA}%Lines break automatically or can be forced with \\
\author{K. Ahmadi-Majlan}
 \affiliation{Department of Physics, The University of Texas at Arlington, 502 Yates Street, Arlington, TX  76019, USA}
\author{X. Shen}%
 \affiliation{Center for Functional Nanomaterials, Brookhaven National Laboratory, Upton, New York 11973, USA}
\affiliation{National Laboratory of Solid State Microstructures and Department of Materials Science and Engineering, Nanjing University, Nanjing 210093, P. R. China}
\author{T. Droubay}%
 \affiliation{Physical Sciences Division, Pacific Northwest National Laboratory, Richland, Washington, USA}
\author{M. Bowden}%
 \affiliation{Environmental Molecular Sciences Laboratory, Pacific Northwest National Laboratory, Richland, Washington, USA}
\author{M. Chrysler}
 \affiliation{Department of Physics, The University of Texas at Arlington, 502 Yates Street, Arlington, TX  76019, USA}
\author{D. Su}%
 \affiliation{Center for Functional Nanomaterials, Brookhaven National Laboratory, Upton, New York 11973, USA}
\author{S. A. Chambers}%
 \affiliation{Physical Sciences Division, Pacific Northwest National Laboratory, Richland, Washington, USA}
\author{J. H. Ngai}
 \affiliation{Department of Physics, The University of Texas at Arlington, 502 Yates Street, Arlington, TX  76019, USA}


\begin{abstract}

The epitaxial growth of crystalline oxides on semiconductors provides a pathway to introduce new functionalities to semiconductor devices. Key to electrically coupling crystalline oxides with semiconductors to realize functional behavior is controlling the manner in which their bands align at interfaces. Here we apply principles of band gap engineering traditionally used at heterojunctions between conventional semiconductors to control the band offset between a single crystalline oxide and a semiconductor. Reactive molecular beam epitaxy is used to realize atomically abrupt and structurally coherent interfaces between SrZr$_{x}$Ti$_{1-x}$O$_3$ and Ge, in which the band-gap of the former is enhanced with Zr content $x$. We present structural and electrical characterization of SrZr$_{x}$Ti$_{1-x}$O$_3$-Ge heterojunctions for $x$ = 0.2 to 0.75 and demonstrate the band offset can be tuned from type-II to type-I, with the latter being verified using photoemission measurements. The type-I band offset provides a platform to integrate the dielectric, ferroelectric and ferromagnetic functionalities of oxides with semiconducting devices.

\end{abstract}

\maketitle 

The integration of multifunctional oxides on conventional semiconductors enables new functionalities to be introduced to semiconductor devices \cite{JWReinerAdvMat}. Perovskite structured oxides (\emph{AB}O$_3$) are of particular interest since a broad range of material behaviors can be realized by simply tuning the stoichiometry of the \emph{A-} and \emph{B-}site ions. The ability to realize a broad range of material behaviors within a single structure is advantageous for epitaxial integration. In this regard, McKee \emph{et al.} pioneered the growth of single crystalline SrTiO$_3$ on Si(100) with atomically abrupt and structurally coherent interfaces\cite{McKeePRL}. Since then, SrTiO$_3$ on Si(100) has been utilized as an epitaxial platform to monolithically integrate multifunctional crystalline oxides on semiconductors\cite{Schlom,DemkovLCO,Ngai}.  
  
Despite the ability to monolithically integrate crystalline oxides on semiconductors, electrically coupling the properties of the former with the latter to realize functional behavior has proven challenging. One of the principal reasons is that SrTiO$_3$ is limited in its ability to act as an \emph{electrical} platform for coupling oxides to semiconductors, despite being an excellent \emph{epitaxial} platform. The band alignment between SrTiO$_3$ and Si, Ge, and GaAs is type-II, namely, the conduction (valence) band of the SrTiO$_3$ is below the conduction (valence) band of the semiconductor \cite{Chambers, FAmy, LiangSTOGaAs}. To couple dielectric, ferroelectric and ferromagnetic properties of oxides with semiconductors, a platform with a type-I band offset in which the conduction (valence) band of the oxide is above (below) the conduction (valence) band of the semiconductor is needed. For example, crystalline SrTiO$_3$ was extensively explored as a gate dielectric on Si due to the low interface trap densities and high dielectric constant that could be achieved through epitaxial growth \cite{WalkerIEEE, Droopad}. However, the type-II band alignment gave rise to excessive leakage currents that limited the effectiveness of crystalline SrTiO$_3$ as a gate dielectric \cite{Chambers,FAmy}. Consequently, the introduction of an intermediary layer of rocksalt structured  $Ae$O ($Ae$ = Sr, Ba) between the Si and SrTiO$_3$ was necessary to reduce charge leakage \cite{McKee_Science, WalkerIEEE}. Similarly, ferroelectric BaTiO$_3$ has been epitaxially grown on Si and Ge to explore the development of a non-volatile, ferroelectric field-effect transistor \cite{Schlom,Ngai}. However, the near alignment of the conduction bands between BaTiO$_3$ and Si or Ge gives rise to leakage currents that prevent the polarization of the former to maintain accumulation or depletion in the latter \cite{Ngai}. 

Thus, controlling band alignments between crystalline oxides and semiconductors is critical to coupling their properties at interfaces. For compound III-V semiconductors, the control of band alignments and band gaps, \emph{i.e.} band-gap engineering, has led to the development of a range of device technologies \cite{Capasso}. Here we apply principles of band-gap engineering, namely, band-gap enhancement through control of stoichiometry, to manipulate the band alignment between a semiconductor and a crystalline perovskite oxide. Reactive oxide molecular beam epitaxy (MBE) is utilized to create heterojunctions between single crystalline SrZr$_x$Ti$_{1-x}$O$_3$ (SZTO) and Ge, in which the band-gap of the former is enhanced through Zr content $x$ \cite{Schafranek}. We show that atomically abrupt and structurally coherent interfaces between SZTO and Ge can be achieved through careful control of kinetic and thermodynamic conditions during deposition. A type-I band offset is manifested through current-voltage ($I$-$V$) and capacitance-voltage measurements ($C$-$V$), in which the former shows significantly reduced gate leakage with Zr content, and the latter indicates that inversion on p-type Ge can be achieved. High-resolution core-level and valence band x-ray photoemission spectroscopy measurements verify and quantify the type-I offset that is achieved for high Zr content $x$. These results demonstrate that band-gap engineering can be exploited to realize functional semiconductor – crystalline oxide heterojunctions. SZTO provides a high-$\kappa$, perovskite structured electrical platform for the integration of dielectric, ferroelectric and ferromagnetic functionalities of oxides onto semiconductors. 

 %%% FIG  %%%%%%%%%%%%%%%%%%%%%%%%%%%%%%
\begin{figure}[t]
\centering
\includegraphics[width=8.5cm]{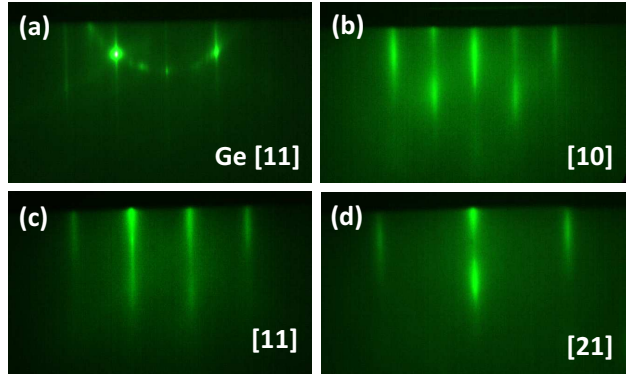}
\caption{\label{Fig1_RHEED} RHEED of SZTO ($x$ = 0.7) on Ge. (a) After 0.5 monolayers of Sr, showing an unusual reconstruction. (b) 38 unit-cell thick SZTO on Ge taken along [10] and (c) [11] and (d) [21] directions.  }
\end{figure} %%%%%%%%%%%%%%%%%%%%%%%%%%% 

Crystalline SZTO films were grown on p-type Ge (100) wafers ($\rho\sim$ 0.02 $\Omega$cm) in a custom MBE chamber. Figure \ref{Fig1_RHEED} shows reflection high energy electron diffraction (RHEED) images obtained at various stages of growth. A half monolayer (ML) of Sr metal was deposited on a clean Ge (100) surface, resulting in the disappearance of the 2 $\times$ structure associated with the dimerized Ge surface and the emergence of a more complicated reconstruction shown in Fig.\ref{Fig1_RHEED} (a). The reconstruction we observe at 0.5 ML Sr coverage is  consistent with the 9 $\times$ structure observed in previous studies \cite{Lukanov}. An initial, 2.5 unit-cell thick layer of amorphous SZTO was deposited at room temperature and subsequently crystallized at high temperature. The perovskite unit-cell of SZTO is rotated 45$^0$ with respect to the diamond-cubic lattice of Ge, $i.e.$ the registry is (001)[100]SZTO//(001)[110]Ge. Additional layers of crystalline SZTO were deposited until 38 unit-cell ($\sim$ 15 nm) thick films were obtained. Figure \ref{Fig1_RHEED} (b), (c) and (d) shows RHEED images of a typical SZTO film along the of the [10], [11] and [21] directions, respectively. A series of films with nominal Zr content ranging from $x$ = 0.2 to 0.75 were grown.

 %%% FIG  %%%%%%%%%%%%%%%%%%%%%%%%%%%%%%
\begin{figure}[t]
\centering
\includegraphics[width=15cm]{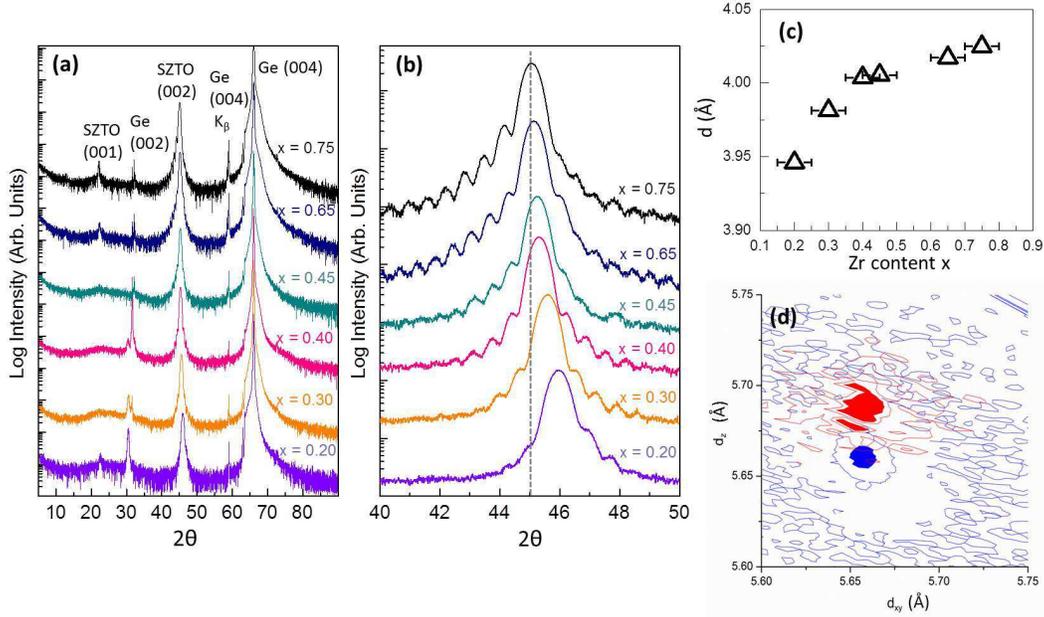}
\caption{\label{Fig2_XRD} (a) Survey scan of SZTO-Ge heterojunctions for various $x$. (b) A shift in the (002) peak is observed with increasing Zr content indicating enhancement of the out-of-plane lattice constant as summarized in (c). (d) Direct-space map of a $x$ = 0.65 SZTO-Ge heterojunction. The lattice constants of SZTO have been multiplied by $\sqrt{2}$ to enable direct comparison with diamond cubic Ge. }
\end{figure} %%%%%%%%%%%%%%%%%%%%%%%%%%%

X-ray diffraction (XRD) measurements confirm single crystalline growth of SZTO on Ge. Figure \ref{Fig2_XRD}(a) shows survey scans of SZTO films of various Zr content. The unit-cell expands along the $c$-axis with increasing Zr, as indicated by a shift in the (002) peak that is shown and summarized in Fig.\ref{Fig2_XRD}(b) and (c), respectively. The surfaces of films of higher Zr content, which are better lattice matched to Ge, are typically smoother than films with less Zr, as evidenced by enhanced finite thickness fringes in the case of the former. Imaging of the films of higher $x$ using atomic force microscopy quantifies the surface roughness to be $<$1 nm \cite{Supplementary}. Similarly, the full-width-at-half-max (FWHM) of rocking curves taken on our films are generally narrower for higher $x$, ranging from $\Delta\omega$$\sim$ 0.39$^0$ to $\sim$ 0.65$^0$. The films remain largely strained to the Ge substrate as shown in the direct space map of Fig.\ref{Fig2_XRD}(d) of a $x$ = 0.65 film. Since the perovskite unit cell of SZTO is rotated 45$^0$ with respect to the diamond-cubic unit cell of Ge, the lattice parameters of the former have been multiplied by $\sqrt{2}$ to enable comparison on the same plot. 
 
%%% FIG  %%%%%%%%%%%%%%%%%%%%%%%%%%%%%%
\begin{figure*}[t!]
\begin{centering}
\includegraphics*[width=14cm]{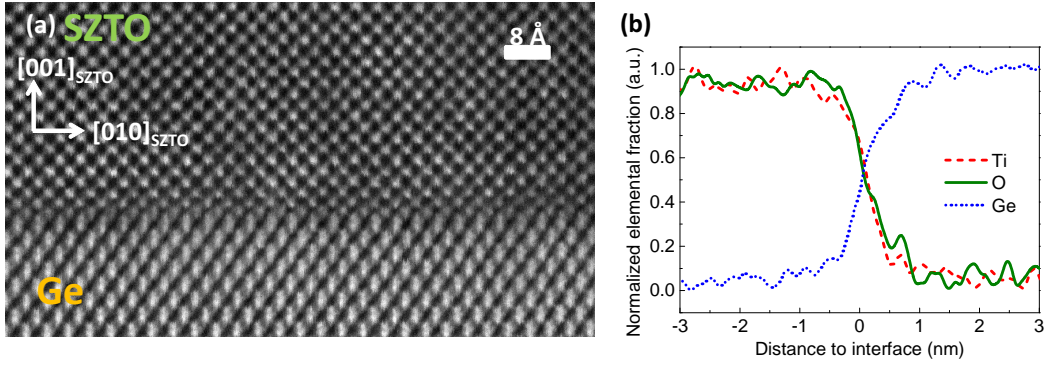}\\
\caption{ (a) STEM characterization of the interface between crystalline SZTO $x$ = 0.70 and Ge (100). The interface between SZTO and Ge is atomically abrupt and structurally coherent. (b) EELS linescan across interface along [001] direction, showing elemental fraction of Ti, O and Ge.  }
\label{Fig3_STEM}
\end{centering}
\end{figure*} %%%%%%%%%%%%%%%%%%%%%%%%%%%%%%%%%%%

Scanning transmission electron microscopy (STEM) measurements provide atomic scale images of the film and interfacial structure between SZTO and Ge. Figure \ref{Fig3_STEM}(a) is a high-angle annular dark field (HAADF) image of a $x$ = 0.70, showing epitaxial registry of the SZTO with the underlying Ge substrate. Particularly noteworthy is the abruptness of the heterojunction, with no extended interfacial layers comprised of amorphous GeO$_2$. The alternating \emph{A}- and \emph{B}-site planes that comprise the perovskite structure is particularly clear due to the closely matched atomic masses of Sr and Zr. Electron energy loss spectroscopy line scans (STEM-EELS) along the [001] direction across the interface corroborate the abruptness of the junction shown in the HAADF image.The distribution of Ti, O and Ge, which are retrieved by integrating the intensity of each EELS spectrum, indicates the thickness of the transition layer is less than 1 nm, as shown in Fig.\ref{Fig3_STEM}(b).

\begin{figure}[t]
\centering
\includegraphics[width=8.5cm]{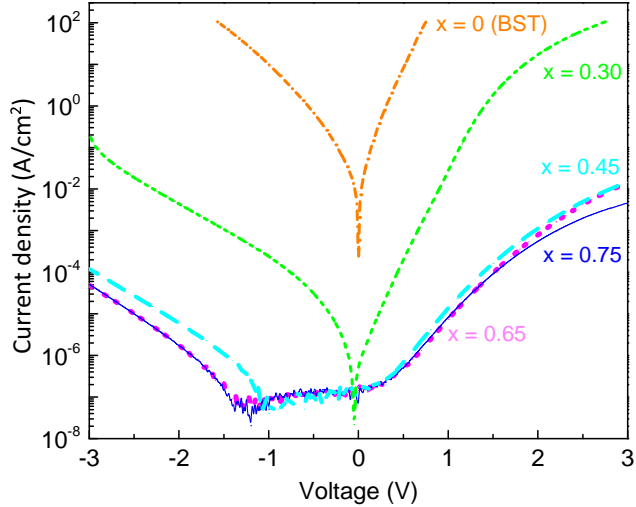}
\caption{\label{Fig4_IV} $I$-$V$ characteristics of SZTO-Ge and BST-Ge heterojunctions. SZTO-Ge heterojunctions exhibit significantly less leakage than BST-Ge heterojunctions, consistent with the presence of a conduction band offset.  }
\end{figure}

The effect of Zr substitution in SZTO on the band alignment at the interface can be seen in measurements of charge leakage through the heterojunctions. Figure \ref{Fig4_IV} shows current-voltage $I$-$V$ measurements through the SZTO-Ge heterojunctions, where the bias is applied to a Ni electrode deposited on top of the SZTO. To demonstrate the effect of Zr substitution in creating a band offset, $I$-$V$ measurements of a 15 nm thick Ba$_{0.4}$Sr$_{0.6}$TiO$_3$(BST) film is also shown. The substitution of Ba for Sr reduces lattice mismatch with Ge but has little effect on the band-gap or offset. The SZTO-Ge heterojunctions exhibit leakage current that is orders of magnitude less than the BST-Ge heterojunction, consistent with the presence of a type-I band offset. An asymmetry in the $I$-$V$ characteristics is observed about zero-bias for SZTO-Ge and BST-Ge heterojunctions, which we attribute to a difference in barrier heights for gate and substrate injection \cite{WalkerIEEE}. Finally, we find that the leakage current has negligible dependence on the area of the junction, indicating a low density of defects throughout the SZTO Ge films.

%%% FIG  %%%%%%%%%%%%%%%%%%%%%%%%%%%%%%
 %%%%%%%%%%%%%%%%%%%%%%%%%%%

\begin{figure}[t]
\centering
\includegraphics[width=8.5cm]{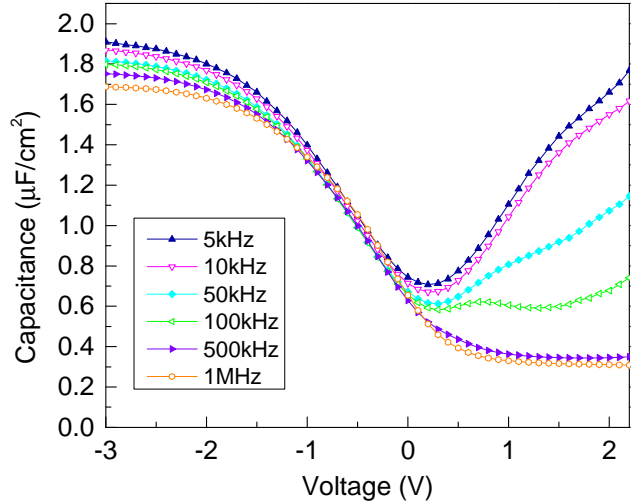}
\caption{\label{Fig5_CV} $C$-$V$ characteristics of SZTO-Ge heterojunctions ($x$ = 0.70). Bias voltage is applied to a Ni electrode.     }
\end{figure}

The type-I band offset is also manifested in capacitance-voltage measurements of our SZTO-Ge heterojunctions. $C$-$V$ measurements of a $x$ = 0.70 SZTO-Ge heterojunction are shown in Fig. \ref{Fig5_CV} as a function of frequency. The capacitance at 1 MHz for $V$ $>$ 1 V matches the expected minimum capacitance for our heavily doped p-type Ge substrates, indicating inversion has been achieved. In the absence of a type-I band offset between the oxide and semiconductor, leakage currents would prevent inversion from occuring on p-type substrates. A shift in the flat-band voltage is observed, which we attribute to positive charges in the SZTO that are likely associated with oxygen vacancies, as has been found in related studies \cite{McKee_Science,WalkerIEEE}.  We estimate the dielectric constant of the SZTO to be $\kappa_{SZTO}$ = 29, based on the capacitance in accumulation at 1 MHz (\emph{i.e.}, 1.7 $\mu$Fcm$^{-2}$). The value of $\kappa_{SZTO}$ = 29, which is near the dielectric constant of bulk SrZrO$_3$ ($\sim$  30), \cite{SZO_kappa} is comparable to some of the highest dielectric constants amongst gate materials that have been reported for Ge \cite{Mavrou}. In this regard, crystalline SZTO on Ge can potentially be utilized as a gate dielectric for next-generation field-effect devices. The performance of complementary-metal-oxide-semiconductor (CMOS) devices has reached a plateau due in part to the fundamental materials limitations of Si. Key to extending CMOS technology is to replace the Si channel with a high mobility material such as Ge \cite{Wallace}. A more thorough evaluation of crystalline SZTO on Ge as a gate dielectric, including thickness dependence of dielectric properties and analysis of interface state densities, will be presented elsewhere \cite{Moghadam}. 

%%% FIG  %%%%%%%%%%%%%%%%%%%%%%%%%%%%%%
 %%%%%%%%%%%%%%%%%%%%%%%%%%%

\begin{figure}[t]
\centering
\includegraphics[width=14cm]{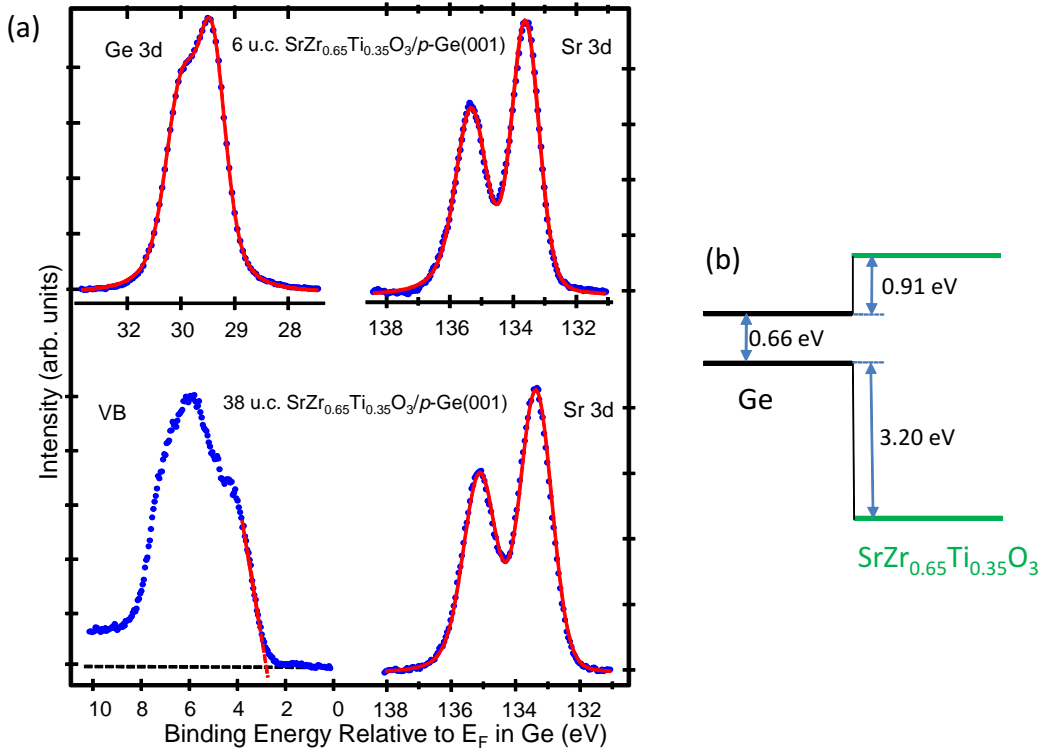}
\caption{\label{Fig6_XPS} (a) Core-level and valence band x-ray photoemission spectra taken on a 6 u.c. thick (top) and 38 u.c. thick (bottom) $x$ = 0.65 SZTO-Ge heterojunctions. Solid lines are fits to the data. (b) Band diagram showing conduction and valence band offsets for a $x$ = 0.65 SZTO-Ge heterojunction.    }
\end{figure}

To quantify the valence band offsets of SZTO on Ge, we have performed high-resolution core-level (CL) and valence band (VB) x-ray photoemission spectroscopy measurements on a $x$ = 0.65 film. We used a combination of CL and VB spectra for thin and thick-film samples to determine the valence band offset (VBO) \cite{KrautPRL,KrautPRB} similar to what was done for SrTiO$_3$/Si(001) \cite{Chambers, ChambersJVAC}. The Sr 3d and Ge 3d peaks were used because they are intense, and do not overlap with XPS and Auger peaks from other elements. The results are shown in Fig.\ref{Fig6_XPS}.  The energy from the top of the VB to the Sr 3$_{d5/2}$ core peak, (E$_{Sr3d5/2}$ $-$ E$_V$)$_{SZTO}$, was measured to be 130.59(6) eV for the 38 u.c. film. This quantity was combined with an analogous energy difference for clean Ge(110), (E$_{Ge3d5/2}$ $-$ E$_V$)$_{Ge}$  = 29.52(4) eV \cite{Waldrop}, and the core-level binding energy difference for the 6 u.c. heterojunction, (E$_{Sr3d5/2}$ $-$ E$_{Ge3d5/2}$)$_{HJ}$  = 104.27(2) eV, to yield a VBO given by $\Delta$E$_V$  = (E$_{Sr3d5/2}$ $-$ E$_{Ge3d5/2}$)$_{HJ}$ + (E$_{Ge3d5/2}$ $-$ E$_V$)$_{Ge}$  $-$ (E$_{Sr3d5/2}$ $-$ E$_V$)$_{STZO}$ = 3.20(8) eV. On the premise the band-gap of SZTO follows Vegard's law, the conduction band offset for $x$ = 0.65 is estimated to be 0.91 eV.  
  
%%% FIG  %%%%%%%%%%%%%%%%%%%%%%%%%%%%%%
 %%%%%%%%%%%%%%%%%%%%%%%%%%%

Aside from serving as a candidate gate dielectric for field-effect devices, SZTO can serve as both an \emph{electrical} and \emph{epitaxial} platform for integrating multifunctional crystalline oxides on semiconductors. For ferroelectrics, SZTO could serve as a intermediate layer to inhibit charge transfer from a ferroelectric to a semiconductor, thus enabling a capacitor to form in which the polarization of the former maintains accumulation or inversion in the latter \cite{Ngai}. As an intermediate buffer, the relatively high-$\kappa$ would reduce effects of depolarization fields in the ferroelectric. For spintronic applications, SZTO could serve as a tunnel barrier through which spin-polarized carriers from a crystalline half-metallic oxide can be injected into a semiconductor. As a candidate tunnel barrier, SZTO is single crystalline, which could potentially enhance injection efficiency \cite{Butler}. Furthermore, the barrier height is adjustable through control of Zr content. 

Previously, rocksalt structured $Ae$O ($Ae$ = Sr, Ba, Ca) were explored as intermediary buffer layers between Si or Ge and $Ae$TiO$_3$ to mitigate the absence of a type-I band offset \cite{McKee_Science}. However, $Ae$O can react with $Ae$TiO$_3$ leading to instability of the rocksalt/perovskite stack \cite{Fompeyrine_stability}. Epitaxially integrating additional layers of multifunctional oxides on $Ae$O is thus challenging. Furthermore, $Ae$O exhibit relatively low dielectric constants, limiting their effectiveness as gate dielectrics. Also, $Ae$O are hygroscopic and are thus unstable in ambient conditions. Finally, the band offset between $Ae$O and semiconductors cannot be continuously tuned from type-II to type-I. Thus, our approach to directly engineer the band-gap of a perovskite oxide on a semiconductor circumvents the challenges of combining rocksalt and perovskite oxides in heterostructures, and enables additional tunability of the band offset. 

Lastly, we make a few remarks on the flexibility of the perovskite structure in enabling band offsets to be engineered at semiconductor - crystalline oxide interfaces. The enhancement of the band-gap and band offset with Zr substitution is accompanied by a concomitant increase in the lattice constant of SZTO, thereby improving the lattice match with Ge, as shown in Fig.\ref{Fig2_XRD}. Generally, it would be more ideal to be able to tune the offset and lattice constant independently. In this regard, the perovskite structure allows for substitution of both \emph{A-} and \emph{B-}site cations. Whereas \emph{B-}site substitution affects both the band-gap and lattice constant of SZTO, \emph{A-}site substitution predominantly affects the lattice constant only. The substitution of Ba (Ca) for Sr could be utilized to increase (decrease) the lattice constant, independent of the Ti to Zr ratio within some regime. Thus, high quality epitaxial films that match well with the semiconducting substrate can, in principle, be achieved for a desired band offset. Future work will focus on exploring this possibility. 
  
In summary, we have applied principles of band-gap engineering to manipulate the band alignment between single crystalline SZTO and Ge, in which the band-gap of the former is enhanced through Zr content. Atomically abrupt and structurally coherent interfaces between SZTO and Ge can be achieved through careful control of kinetic and thermodynamic conditions during deposition. Current-voltage measurements show a significant decrease in gate leakage with Zr content, and capacitance-voltage measurements indicate that  inversion can be achieved. These results demonstrate that band-gap engineering can be exploited to realize functional semiconductor – crystalline oxide heterojunctions. SZTO can serve as a high-$\kappa$, perovskite structured electrical platform for integrating dielectric, ferroelectric and ferromagnetic functionalities of oxides onto semiconductors. The ability to electrically couple multifunctional oxides with semiconductors would enable a range of device technologies to be realized, advancing the oxide - semiconductor interface well beyond the paradigm of the metal-oxide-semiconductor capacitor.   
    
\section{\label{sec:level1}Experimental Section}

Crystalline SZTO films were grown on p-type Ge (100) wafers (AXT Inc., $\rho\sim$ 0.02 $\Omega$cm) using reactive MBE in a custom-built chamber operating at a base pressure of $<$ 2 $\times$ 10$^{-10}$ Torr. The Ge wafers were introduced into the growth chamber after an etch and oxidation process that involved repeated dips in diluted HCl and H$_2$O$_2$\cite{Demkov_Ge}. A clean dimerized Ge surface was obtained by thermally desorbing the resulting GeO$_2$ from the surface in ultra-high vacuum at $\sim$ 600 $^0$C. The Ge was then cooled to $\sim$ 400 $^0$C at which a half monolayer of Sr metal was deposited to passivate the clean Ge surface. Thermal effusion cells (Veeco and SVT Associates) were used to evaporate all source materials for growth, and fluxes were calibrated using a quartz crystal monitor (Inficon).The Ge was then cooled to room temperature at which 1.5 monolayers of SrO was co-deposited with 2 monolayers of Zr$_{x}$Ti$_{1-x}$O$_2$ in an oxygen background pressure of 3 $\times$ 10$^{-7}$ Torr. The Ge was then heated to $\sim$ 550 $^0$C to crystallize the 2.5 unit cells of SZTO. Subsequent SZTO layers were deposited 3 unit cells at a time at a substrate temperature of $\sim$ 250 $^0$C, and background oxygen pressure of 3 $\times$ 10$^{-7}$ Torr. The substrate temperature was increased to $\sim$ 580 $^0$C in ultra-high-vacuum after each 3 unit cell deposition of SZTO to improve crystallinity.

The cross-sectional STEM specimens along [110] direction of Ge were prepared by a focused ion beam (FIB) lift-out technique using a FEI-Helios FIB system. A 5 KV Ga-ion beam was used for final thinning to remove the damaged surface layer. Imaging of the SZTO/Ge heterojunction was performed in an aberration-corrected Hitachi HD2700C dedicated STEM equipped with a high-resolution parallel EELS detector (Gatan Enfina-ER). During the measurement, we used a 1.3 \AA  probe with a beam current of 12 pA. The energy resolution of EELS spectrum was about 0.35 eV. 

XRD measurements were performed in a Bruker D8 thin-film diffractometer using Cu K$\alpha$ radiation. The direct-space map was measured on a Panalytical Materials Research Diffractometer. The Ge (224) and perovskite (103) peaks were mapped separately because they are widely separated in diffraction space. These peaks were chosen because they can both be measured at the same $\phi$ angle (rotation about the sample surface normal) and therefore errors arising from a difference between the [001] and $\phi$-axis are avoided. The scattering vector in real space was resolved into in-plane and out-of-plane components and multiplied by $\sqrt{h^2 + k^2}$ and $l$ respectively to give dimensions matching the lattice parameters. The coordinates of the perovskite SZTO map was further multiplied by $\sqrt{2}$ to account for the 45$^0$ degree epitaxy to allow direct comparison on the same plot. 

High-resolution core-level (CL) and valence band (VB) x-ray photoelectron spectra were measured for as-received SrZr$_{x}$Ti$_{1-x}$O$_3$ $x$ = 0.65 films of thickness equal to 6 u.c. and 38 u.c. using monochromatic Al K$\alpha$ x-rays (h$\nu$ = 1486.6 eV) and a VG/Scienta SES 3000 electron energy analyzer. All spectra were measured in the normal-emission geometry and with a total energy resolution of 0.5 eV. 

Prior to $C$-$V$ and $I$-$V$ characterization, both SZTO-Ge and BST-Ge heterojunctions were annealed for 30 minutes at 350 $^0$C in flowing wet oxygen \cite{wet_oxygen}. As was found in previous studies \cite{Reiner_LAOSi}, low temperature annealing in wet oxygen reduces residual oxygen vacancies. However, this low temperature anneal does not affect the structurally coherent atomic registry at the SZTO and Ge interface ($i.e.$ no GeO$_2$ formed), as revealed in high resolution transmission electron microscopy images of post-annealed samples \cite{Supplementary}. 30 nm thick Ni electrodes ranging in diameter from 20 $\mu$m to 270 $\mu$m were deposited through a shadow mask using electron beam evaporation. The backsides of the Ge wafers were mechanically scratched and InGa eutectic was applied to form a counter electrode. The $I$-$V$ and $C$-$V$ characterizations were performed on a Micromanipulator 8060 probe station using flexible, 10 micron radii, tungsten probe whiskers. An Agilent 4155C and an Agilent 4284A LCR meter were used for the $I$-$V$ and $C$-$V$  measurements, respectively. $I$-$V$ measurements were performed by ramping the voltage applied to the Ni electrode from 0 to +3 V. A different, junction was used to measure leakage current in the range of 0 to -3 V. The $I$-$V$ characteristics summarized in Fig.\ref{Fig4_IV} were averged over 10 junctions for each film.    
		
\begin{acknowledgements}

This work was supported by the University of Texas at Arlington. XS thanks the China Scholarship Council and Brookhaven National Laboratory for support. The STEM characterization performed at the Center for Functional Nanomaterials at BNL was supported by U.S. Department of Energy, Office of Basic Energy Sciences, under contract no. DEAC02-98CH10886. The band offset measurements and the reciprocal space mapping were performed at PNNL and supported by the U.S. Department of Energy, Office of Science, Division of Materials Sciences and Engineering under Award 10122. The PNNL work was performed in the Environmental Molecular Sciences Laboratory, a national science user facility sponsored by the Department of Energy$'$s Office of Biological and Environmental Research. 

\end{acknowledgements}


\begin{references}
\bibitem{wet_oxygen} Wet oxygen is ultra-high purity oxygen that has been bubbled through de-ionized water prior to flowing through a tube furnace.
\end{references}
\end{document}